\documentclass[sigconf,prologue,table]{acmart}

\AtBeginDocument{%
  \providecommand\BibTeX{{%
    \normalfont B\kern-0.5em{\scshape i\kern-0.25em b}\kern-0.8em\TeX}}}

\usepackage{booktabs}
\usepackage{enumitem}
\usepackage{caption}
\usepackage{subcaption}
\usepackage{lineno,hyperref}
\usepackage{multirow}
\usepackage{graphicx}
\usepackage{balance}
\usepackage[linesnumbered,boxruled]{algorithm2e}
\usepackage{algorithm2e}
\usepackage{comment}
\usepackage{array}
\usepackage{algorithmic}
\modulolinenumbers[5]
\usepackage[most]{tcolorbox}

\tcbset{textmarker/.style={
        enhanced,
        parbox=false,boxrule=0mm,boxsep=0mm,arc=0mm,
        outer arc=0mm,left=5mm,right=3mm,top=7pt,bottom=7pt,
        toptitle=1mm,bottomtitle=1mm,oversize}}

\newtcolorbox{hintBox}{textmarker,
    borderline west={6pt}{0pt}{yellow},
    colback=yellow!10!white}
\newtcolorbox{importantBox}{textmarker,
    borderline west={6pt}{0pt}{red},
    colback=red!10!white}
\newtcolorbox{noteBox}{textmarker,
    borderline west={8pt}{0pt}{gray},
    colback=gray!10!white}

\usepackage{pgfplotstable}

\def\BibTeX{{\rm B\kern-.05em{\sc i\kern-.025em b}\kern-.08em
    T\kern-.1667em\lower.7ex\hbox{E}\kern-.125emX}}
\graphicspath{{pics/}}

\newcommand{\Cross}{cross-ecosystem library}
\newcommand{\Crosses}{cross-ecosystem libraries}

\newcommand{\RqOne}{\textbf{(RQ2)} \emph{What percent of contributors to a \Cross~repository are from different ecosystems?}}

\newcommand{\RqOneR}{
A majority (median of 37.5\%) of contributors come from a single ecosystem, while a  significant portion of contributors do not belong
to any of those two ecosystems (median of 24.06\%).
}

\newcommand{\PQ}{\textbf{(RQ1)} \emph{How dependent is the ecosystem to a \Cross~release?}}

\definecolor{chestnut}{rgb}{0.8, 0.36, 0.36}

\acmConference[MSR 2023]{MSR '23: Proceedings of the 20th International Conference on Mining Software Repositories}{May 15-16, 2023}{Melbourne, Australia}

\begin{document}
\begin{sloppy}

\title{Intertwining Communities: \\Exploring Libraries that Cross Software Ecosystems}

\author{Kanchanok Kannee, Raula Gaikovina Kula, Supatsara Wattanakriengkrai, Kenichi Matsumoto \\
Nara Institute of Science and Technology, Japan\\
Email: kanchanokkannee@gmail.com,
\{raula-k, wattanakri.supatsara.ws3, matumoto\}@is.naist.jp}

\begin{abstract}
Using libraries in applications has helped developers reduce the costs of reinventing already existing code. However, an increase in diverse technology stacks and third-party library usage has led developers to inevitably switch technologies and search for similar libraries implemented in the new technology. To assist with searching for these replacement libraries, maintainers have started to release their libraries to multiple ecosystems. Our goal is to explore the extent to which these libraries are intertwined between ecosystems. We perform a large-scale empirical study of 1.1 million libraries from five different software ecosystems, i.e., PyPI, CRAN, Maven, RubyGems, and NPM, to identify 4,146 GitHub repositories. As a starting point, insights from the study raise implications for library maintainers, users, contributors, and researchers into understanding how these different ecosystems are becoming more intertwined with each other.
\end{abstract}
\maketitle


\section{Introduction}
\label{sec:introduction}
Popular use of third-party libraries has become prominent in contemporary software engineering \cite{KulaEMSE2018}, which is evident by the emergence of different library repositories like NPM, PyPI, CRAN, Maven, and so on.
These massive repositories also depend on each other, thus forming a complex software ecosystem of dependencies.

For various reasons, a developer may realize that a library used in their applications requires replacement, either due to the availability of newer versions that fix defects, patch vulnerabilities, and enhance features.
In such cases, the developer seeks an appropriate replacement and has led to various efforts in library recommendation \cite{Cossette2021recommendation}.
However, when faced to switch programming language, the options may become limited. 
With an increase of the diversity in technology stacks and third-party library usage \cite{Syful2021ICSME}, developers eventually will face the need to switch a programming language and their subsequent libraries specific to that language \cite{analogicalqa,Chen2016QA,Teyton2013map}.

\begin{figure*}[]
    \centering
    \begin{subfigure}{0.5\linewidth}
         \includegraphics[width=\textwidth]{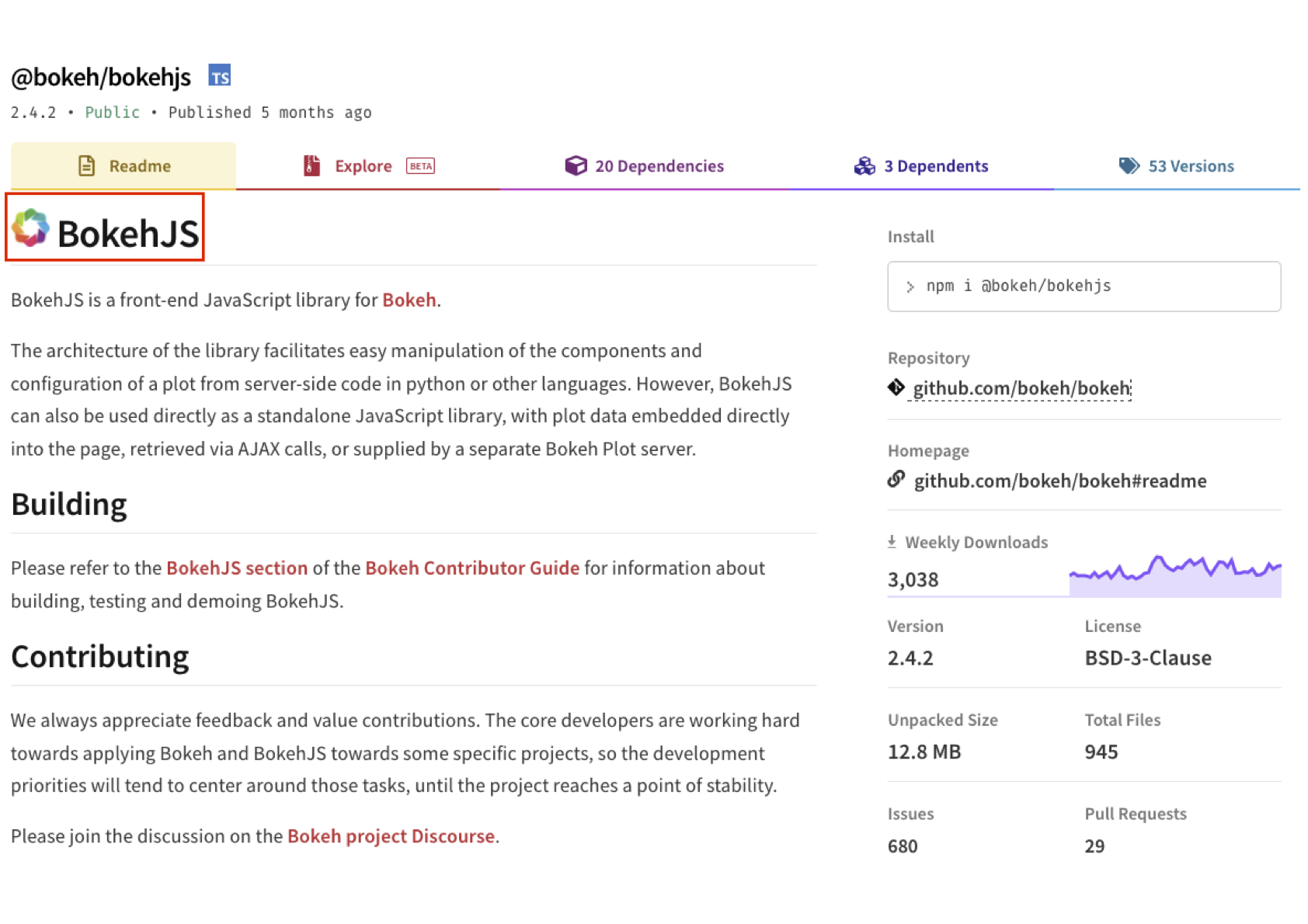}
         \caption{@bokeh/bokehjs}
     \end{subfigure}
     \begin{subfigure}{0.45\linewidth}
         \includegraphics[width=\textwidth]{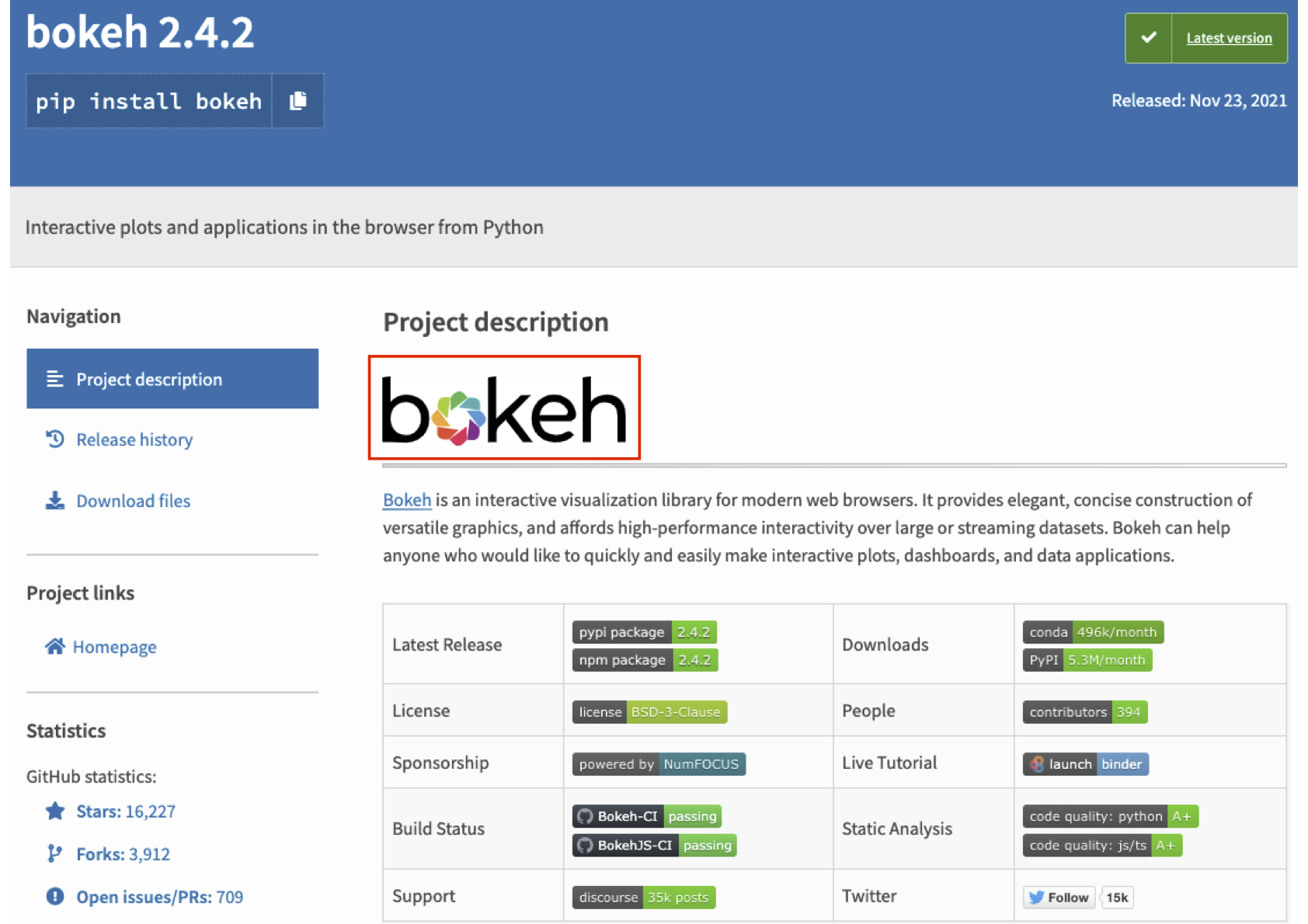}
         \caption{bokeh}
     \end{subfigure}
    \caption{Our example of the Bokeh library, released to the PyPI (Bokeh) and NPM (BokehJS) ecosystems}
    \label{fig:bokeh}
\end{figure*}

To facilitate the movement from one programming language to another, a library may release a version that is specific for that library ecosystem, thus the library maintainers decide to create a library that \textit{intertwines two or more ecosystems}.
Figure~\ref{fig:bokeh} depicts a library that serves two different ecosystems (cf. Figure~\ref{fig:bokeh}a and Figure~\ref{fig:bokeh}b).
The library is Bokeh \cite{bokehbok12:online},
which is a popular interactive visualization library for modern web browsers.
The official GitHub repository that hosts the common repository, while there are two versions of the library hosted on both the official PyPI\cite{bokehPy80:online} and NPM registry \cite{bokehbok61:online}.
By 2022, the Python release of Bokeh had over 2.94K dependent repositories, and 590 other libraries in the ecosystem that rely on this library.
It has 120 releases and was first released on October 25th, 2013.
The GitHub repository that hosts the \Cross~is mainly implemented in Python (i.e., 56.7\%) and 
TypeScript (i.e., 41.1\%).
According to its homepage, BokehJS is written primarily in TypeScript and in JavaScript.
Furthermore, Bokeh has attracted 603 contributors to its GitHub repository.

In this short paper, we conduct a large scale quantitative study to explore the extent to which these \Crosses~are intertwined within these ecosystems.
Since prior work \cite{fse2018_sustained} shows that the ecosystem plays an important role in the sustained activities of a software project, we explore the extent to which these libraries may require involvement from multiple ecosystems.
Our study complements the research conducted by Constantinou et al. \cite{Constantinou2018BreakingTB} which investigated the presence and characteristics of \Crosses~in twelve software distributions.
We mine five of the most popular and widely adopted library ecosystems (i.e., CRAN, Maven, PyPI, RubyGems, and NPM).
Our study is a large-scale quantitative analysis of 1,110,059 libraries to identify 4,146 GitHub repositories with 567,864 contributors, asking two research questions:

\begin{itemize}
\item \noindent \PQ~\\
\textit{Results:} We find that cross-ecosystem libraries belong to four out of the seven ecosystem pairs are depended upon (i.e., NPM, Maven, PyPI and RubyGems).
\item \noindent \RqOne~\\
\textit{Results:} \RqOneR 
\end{itemize}

The results reveal that communities do reach beyond the boundaries of a single programming language.
We make our dataset available which is a large quantitative study that covers over 1.1 million libraries, and 500 thousand contributors.
The second contribution is a replication package with all data and scripts available at \url{https://doi.org/10.5281/zenodo.6524901}.

\section{Data Preparation}

\paragraph{Target Package Ecosystems}
We selected five popular and well-studied software ecosystems \cite{10.1145/2993412.3003382, Cesar:emse2021, decan2019empirical}. 
NPM is a package manager for the JavaScript programming language that was recently purchased by Microsoft via GitHub on March 16, 2020\cite{npm27:online}.
PyPI is the library ecosystem that serves the Python programming language, which is interpreted as a high-level general-purpose programming language.
CRAN is the library ecosystem that serves the R programming language, which is a free software environment for statistical computing and graphics \cite{RTheRPro33:online}.
Maven is the library ecosystem that serves the Java programming language,  which is a general-purpose programming language that follows the object-oriented programming paradigm and can be used for desktop, web, mobile, and enterprise applications\cite{JavaOrac42:online}.
RubyGems is the library ecosystem that serves the Ruby programming language, which is a dynamic, open-source programming functional programming language with a focus on simplicity and productivity\cite{RubyProg16:online}.

\begin{table}[]
\centering
\caption{Overview of our dataset}
\label{tab:collected_repo}
\begin{tabular}{lcc}
\hline
\multicolumn{1}{c|}{\multirow{2}{*}{\# Lib. Releases}} & \multicolumn{2}{c}{As of 12 Jan 2020} \\ \cline{2-3} 
\multicolumn{1}{c|}{}                                             & with GitHub Repo URL    & initial libraries  \\ \hline
\multicolumn{1}{l|}{ NPM}                                & 818,787                 & 2,357,829       \\
\multicolumn{1}{l|}{ PyPI}                                    & 138,001                 & 420,350        \\
\multicolumn{1}{l|}{ CRAN}                                    & 5,551                   & 21,526       \\
\multicolumn{1}{l|}{ Maven}                                   & 36,762                  & 456,756   \\
\multicolumn{1}{l|}{ RubyGems}                                & 110,958                 & 176,987    \\ \hline
\multicolumn{1}{c|}{Total}                                & 1,110,023                & 3,433,448   \\
\multicolumn{1}{c|}{\# pairs}                       & \multicolumn{2}{c}{4,146}   \\\hline
\end{tabular}
\end{table}

\begin{figure}[]
    \centering
    \frame{\includegraphics[width=9cm]{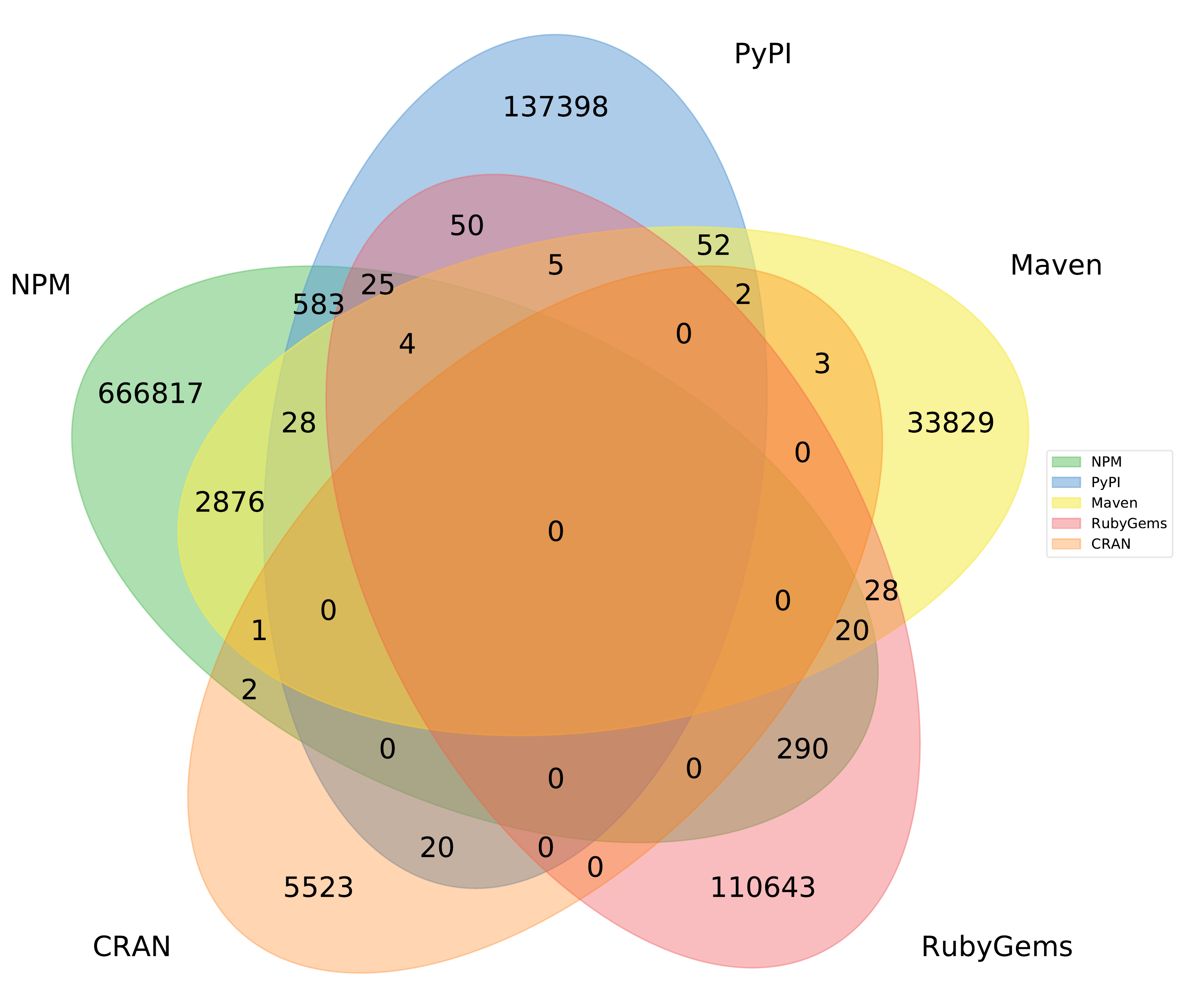}}
    \caption{4,146 libraries that cross five ecosystems. }
    \label{fig:star_data}
\end{figure}

\begin{table}[]
\centering
\caption{Contributors from the 1,110,023 libraries}
\label{tab:data_for_each_rq}
\begin{tabular}{l|ccr}

\multicolumn{1}{l}{} &  As of            &  March 2022            &                \\
\cline{2-4}
                                               & {Median} & {Max} & {Total} \\ \hline
\multicolumn{1}{l|}{\textbf{\# Contrib.}} &                 &              &                \\
\multicolumn{1}{r|}{per paired library.}                  & 8               & 1,727        & 49,674         \\
\multicolumn{1}{r|}{per lib.}     & 4               & 16,606       & 567,864        \\ \hline
\end{tabular}
\end{table}

\paragraph{Detection Method}
To identify a library that crosses multiple ecosystems, we use the GitHub library repository as the linking heuristic.
For instance, as shown in our motivating example, the GitHub library repository URL serves as a link between two analogical libraries.
Our assumption is that GitHub should be the common platform by which these libraries host their source code.
We also use this filter as a quality sanity check.
Similar to prior work \cite{zerouali2019diversity,decan2019empirical,alfadel2021use}, we then queried the Libraries.io dataset for library ecosystems that indeed listed a GitHub repository URL as their library repository.
For each library ecosystem, we collect a list of all libraries from the Libraries.io dataset\cite{Librarie69:online}.
We mined the dataset version (1.6.0)\cite{jeremy_katz_2020_3626071}.
Once we were able to collect the list of libraries that hosted their library repository on GitHub, we then proceed to cross-reference between libraries that are hosted in different library ecosystems.

Table \ref{tab:collected_repo} shows a summary of the extracted 1,110,023 GitHub repositories, while Figure \ref{fig:star_data} depicts GitHub repositories belonging to libraries that are released to multiple ecosystems. 
We notice from this figure that these libraries are usually comprised of pairs (i.e., NPM $\cap$ PyPI).

Table \ref{tab:data_for_each_rq} shows the summary statistics of contributors. We then mined to answer RQ2. We use the GitHub API\cite{GitHubRE50:online} to collect all contributors that made commits to all repositories. 
We use the API request https://api.github.com/repos/$\{$owner$\}$/$\{$repo$\}$/commits to collect all the commit information.
After collecting all contributors for each library, we then merged the contributors' list based on the ecosystem, so that we have a merged list of contributors.
We classified a contributor as belonging to an ecosystem if they made commits to at least two or more libraries.
Due to compliance with GitHub API terms of usage, we slowly downloaded this information for two months to collect all contributor information.

\section{Dependence on the Ecosystem (RQ1) }
\label{sec:Findings}

\begin{table}[]
\centering
\caption{Summary statistics of the \# of Dependents, showing statistical differences including significance (Sig.) the libraries with multiple ecosystems (Eco. Pairs) are more dependent than regular libraries.}
\label{tab:preliminary}
\begin{tabular}{@{}crrrrc@{}}
\toprule
\multicolumn{1}{l}{} & \multicolumn{4}{c}{\# Dependents} & \multicolumn{1}{l}{} \\
\multicolumn{1}{l}{} & \multicolumn{2}{c}{Mean} & \multicolumn{2}{c}{Median} & \multicolumn{1}{l}{} \\ \midrule
Ecosystem & \multicolumn{1}{c}{Eco. Pairs} & \multicolumn{1}{c}{Regular} & \multicolumn{1}{c}{Eco. Pair} & \multicolumn{1}{c}{Regular} & \multicolumn{1}{l}{Sig.} \\ \midrule
 & \cellcolor[HTML]{9AFF99}2664.71 & \cellcolor[HTML]{9AFF99}761.39 & \cellcolor[HTML]{9AFF99}53 & \cellcolor[HTML]{9AFF99}1 & \cellcolor[HTML]{9AFF99}* S \\
\multirow{-2}{*}{\begin{tabular}[c]{@{}c@{}}NPM  \\ PyPI\end{tabular}} & 405.35 & 29.43 & 26 & 0 & - \\ \midrule
 & 2884.94 & 761.39 & 71 & 1 & - \\
\multirow{-2}{*}{\begin{tabular}[c]{@{}c@{}}NPM \\ Maven\end{tabular}} & \cellcolor[HTML]{9AFF99}75.06 & \cellcolor[HTML]{9AFF99}59.35 & \cellcolor[HTML]{9AFF99}5 & \cellcolor[HTML]{9AFF99}0 & \cellcolor[HTML]{9AFF99}* N \\ \midrule
 & 4003.08 & 761.39 & 19 & 1 & - \\
\multirow{-2}{*}{\begin{tabular}[c]{@{}c@{}}NPM \\ RubyGems\end{tabular}} & 2291.94 & 335.90 & 16 & 0 & - \\ \midrule
 & 21.43 & 19 & 12 & 0 & - \\
\multirow{-2}{*}{\begin{tabular}[c]{@{}c@{}}CRAN  \\ PyPI\end{tabular}} & - & - & - & - & - \\ \midrule
 & 94.21 & 59.35 & 20 & 0 & - \\
\multirow{-2}{*}{\begin{tabular}[c]{@{}c@{}}Maven  \\ PyPI\end{tabular}} & 91.42 & 29.43 & 23 & 0 & - \\ \midrule
 & \cellcolor[HTML]{9AFF99}269.47 & \cellcolor[HTML]{9AFF99}59.35 & \cellcolor[HTML]{9AFF99}34 & \cellcolor[HTML]{9AFF99}0 & \cellcolor[HTML]{9AFF99}* S \\
\multirow{-2}{*}{\begin{tabular}[c]{@{}c@{}}Maven  \\ RubyGems\end{tabular}} & \cellcolor[HTML]{9AFF99}1104.91 & \cellcolor[HTML]{9AFF99}335.90 & \cellcolor[HTML]{9AFF99}56 & \cellcolor[HTML]{9AFF99}0 & \cellcolor[HTML]{9AFF99}* S \\ \midrule
 & \cellcolor[HTML]{9AFF99}132.66 & \cellcolor[HTML]{9AFF99}29.43 & \cellcolor[HTML]{9AFF99}7 & \cellcolor[HTML]{9AFF99}0 & \cellcolor[HTML]{9AFF99}* S \\
\multirow{-2}{*}{\begin{tabular}[c]{@{}c@{}}PyPI  \\ RubyGems\end{tabular}} & \cellcolor[HTML]{9AFF99}569.63 & \cellcolor[HTML]{9AFF99}335.90 & \cellcolor[HTML]{9AFF99}8 & \cellcolor[HTML]{9AFF99}0 & \cellcolor[HTML]{9AFF99}* S \\ \bottomrule
\multicolumn{5}{l}{\begin{tabular}[c]{@{}l@{}}The effect sizes level: N(negligible), and S(small)\\ *:p-value 	$<$ 0.05\end{tabular}} 
\end{tabular}
\end{table}

\paragraph{\textit{Approach}}

To answer RQ1, we collected library dependents, which is the number of other libraries in the same ecosystem that declares this library as a dependency.
Note that for the analysis, we compare against a pair of ecosystems (e.g., NPM $\cap$ PyPI), resulting in two dependency scores.
For example, Font-Awesome\cite{FortAwes12:online} is one of the top open-source libraries on GitHub, used by millions of designers, developers, and content creators. In this case, the library has a dependency score for both the PyPI (i.e., 29 dependents) and Maven (i.e., 22 dependents).
For evaluation, we compare the libraries that pair different ecosystems to a baseline (i.e., other regular libraries) and report the statistical summary (i.e., mean, median).

To statistically validate our results, we use the Mann-Whitney U test~\cite{mann1947test}, \cite{wilcoxon1945individual}, which is a non-parametric statistical test.
To show the power of differences between metrics from cross-ecosystem libraries and regular libraries (i.e., libraries belong to one ecosystem), we investigate the effect size using Cliff’s $\delta$, which is a non-parametric effect size measure \cite{romano2006exploring}. 
The interpretation of Cliff's $\delta$ is shown as follows: (1) $\delta$ $<$ 0.147 as Negligible,
(2) 0.147 $\leq$  $\delta$  $<$ 0.33 as Small, (3) 0.33 $\leq$  $\delta$  $<$ 0.474 as Medium, or (4)
0.474 $\leq$  $\delta$  as Large. To analyze Cliff’s $\delta$, we use the cliffsDelta package\cite{neilerns80:online}.

\paragraph{\textit{Results}}
Table \ref{tab:preliminary}
shows the evidence that the cross-ecosystem library is highly dependent upon by the multiple ecosystems.
Results show certain pairings with NPM and RubyGems have a large pairing.
Statistically, we find that the number of dependents against the baseline of regular libraries are significantly different (p-value $<$ 0.05), with a negligible to small association as the reported effect size for different combinations.
We note that negligible and small effect sizes are noticeably smaller but not so small as to be trivial according to statistical analysis.
Further research needs to examine how various factors, such as the size of the ecosystem, contribute to the small or negligible statistical variances.

\begin{tcolorbox}
\textbf{Summary for RQ1:} 
Four out of the seven ecosystem pairs of libraries depended on cross ecosystem libraries.
(i.e., NPM, Maven, PyPI and RubyGems)
\end{tcolorbox}

\section{Distribution of Contributors (RQ2)}

\paragraph{Approach}

To answer RQ2, we investigate whether a \Cross~attracts contributions from the targeted ecosystems.
For data collection, we first collected all contributors from the 1,110,059 projects.
For our analysis, we calculated a percentage of three types of contributors of each \Cross:
\begin{itemize}
    \item \texttt{Both (\%)} refers to the percentage of contributors that have made prior contributions to libraries that belong to both ecosystems.
    \item \texttt{Single (\%)} refers to the percentage of contributors that have made prior contributions to a single ecosystem.
    \item \texttt{Independent (\%)} refers to the percentage of contributors who do not have any contributions to any ecosystem libraries.
\end{itemize}

Since we are dealing with a large dataset, we used the PySpark\cite{PySparkD87:online} arrays to handle the data processing tasks of classifying and and merging contributors into these three groups.
To statistically confirm our classification of contributors, we use McNemar’s Chi-Square test \cite{McNemar1947}. This is a non-parametric statistical test used to find the change in proportion for paired data. 
We test the null hypothesis that \textit{`the percentage of contributions classified is the same'}
We also measure the effect size using Cohen’s $d$, a non-parametric effect size measure by \cite{cohen2013statistical}. Effect size is analyzed as follows: 
(1) $ d < 0.2$ as Negligible, (2) $0.2 \leq d <0.5$ as Small, (3) $0.5 \leq d <0.8$ as Medium, or  (4) $0.8 \leq d$ as Large.

\paragraph{Results}

From Figure \ref{fig:RQ1_violinplot} we make two observations. 
First, we find that there is a higher percentage of contribution (a median of 37.5\%) that is supported by only one ecosystem. (see Figure~\ref{fig:RQ1_violinplot}).
It is important to note that our analysis does not identify the original ecosystem.
The second observation is that a significant portion of contributors  (median of 24.06\%) that do not belong to either ecosystem.
The result shows that these contributors may be just specific to the project itself. 
This is more than contributors that belong to both ecosystems (median of 20\%).

 Table~\ref{tab:sta_RQ1} shows statistical comparison between groups of contributors. We find significant differences between (p-value $<$ 0.05) where a library that has a higher percentage of single ecosystem contributors will have a lower percentage of contributors from both ecosystems.

\begin{figure}[]
    \centering
    \frame{\includegraphics[width=6cm]{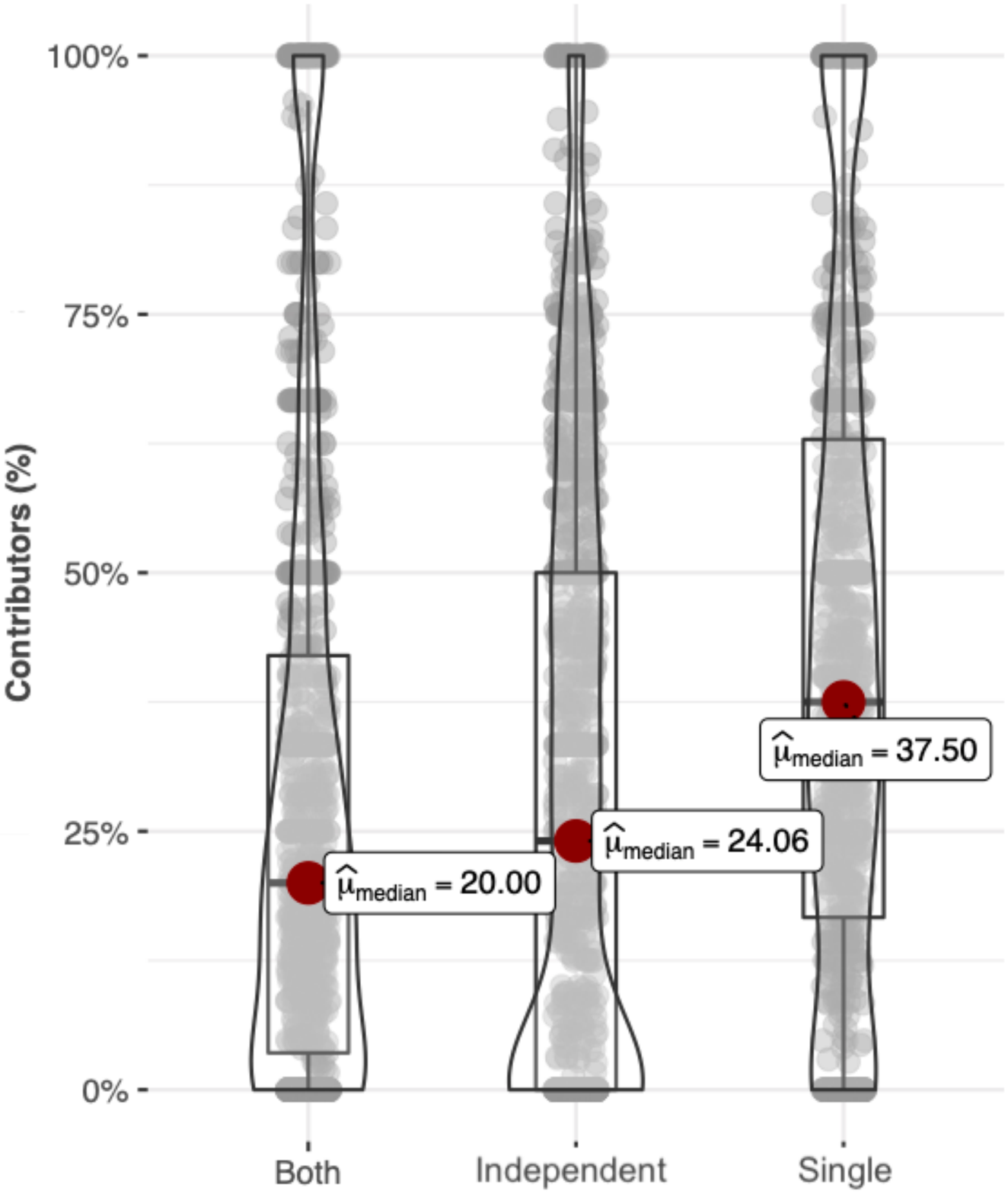}}
    \caption{Answering RQ2, we show how contributions for a \Cross~originate from a single ecosystem.}
    \label{fig:RQ1_violinplot}
\end{figure}

\begin{table}[]
\centering
\caption{Statistical significance test results related to RQ2. }
\label{tab:sta_RQ1}
\begin{tabular}{cc}
\hline
\multicolumn{2}{|c|}{\textbf{Contributions (\%)}}                                                                          \\ \hline
\multicolumn{1}{|c}{{Single-Both *}}                        & \multicolumn{1}{c|}{S}                     \\
\multicolumn{1}{|c}{{-}}                     & \multicolumn{1}{c|}{-}                     \\ 
\multicolumn{1}{|c}{Both-Neither *}                     & \multicolumn{1}{c|}{N}                     \\ \hline
\multicolumn{2}{l}{\begin{tabular}[c]{@{}l@{}}The effect sizes level: small(S) and negligible(N)\\ *:p-value \textless 0.05\end{tabular}}
\end{tabular}
\end{table}

\begin{tcolorbox}
\textbf{Summary for RQ2:} \RqOneR
\end{tcolorbox}

\section{Limitations}
\label{sec:threats_to_validity}

A key threat in the construct validity exists in the matching approach using the GitHub URL, which reduces our study to GitHub projects. 
Also, results are  limited to the five ecosystems: NPM, PyPI, CRAN, Maven, and RubyGems, which threatens the generalization of our claims to other ecosystems, and is seen as immediate future work.

\section{Research Agenda}
\label{sec:recommendations}
As an alternative to replacement libraries, our results show promising results on the phenomenon of releasing a library to multiple ecosystems. We summarize three main questions below:

\paragraph{\textbf{Should maintainers consider releasing to \\ multiple ecosystems?}}
Results from RQ1 highlight that not only this phenomenon exists, but that in some ecosystems that library is dependent upon by the ecosystem.
This does raise the question of whether or not maintainers should consider following suit.
However, as shown in RQ2, current contributions are not shared by both, but only a single ecosystem.
Results from the study also raise questions on how developers may need to be proficient in multiple programming languages?
For future work, there needs to be a qualitative analysis or case studies to gain a deeper understanding about what are the concrete benefits of opening up to a new ecosystem (e.g., attracting contributors, increasing code quality and finding bugs), and motivations for releasing to multiple ecosystems. 
This could be done by either mining the library repositories, or by survey interviews. 

\paragraph{\textbf{Will this phenomenon solve the need to find replacement libraries?}}
From the results of RQ1, we cannot conclude the answer to this question. 
However, the results from RQ1 show that the more dependent and somewhat mature libraries are taking up this trend. 
Another interesting research direction is to see the impact of these libraries on the existing libraries that already provide these functionalities.
An engaging avenue is how these libraries will compete with already established libraries in the new ecosystem that provide the same functionalities.
This may be at the finer grain of certain libraries, certain ecosystems, and user surveys of maintainers. 

\paragraph{\textbf{How will cross-ecosystem libraries impact ecosystem-level topics like governance, and management?}}
From a research perspective, the growing intertwining between different ecosystems will bring forth interesting implications at the ecosystem level.
From one point of view, the results indicate that the boundaries of a community are not limited by the programming language.
However, with this expansion, the extent of governance and management is unknown. 
For instance, to what extent do these libraries abide by the specific rules and regulations that are enforced by each ecosystem? And how are bug fixes and specific security vulnerabilities propagated through different ecosystems?
Since RQ2 states that their libraries are more likely to receive contributions from the original ecosystem, does that mean that the boundaries between these two ecosystems become closer? 

\section*{Acknowledgement}
This work is supported by Japanese Society for the Promotion of Science (JSPS) KAKENHI Grant Numbers 20K19774 and 20H05706.

\bibliographystyle{ACM-Reference-Format}
\bibliography{bibliography}


\begin{thebibliography}{31}


\ifx \showCODEN    \undefined \def \showCODEN     #1{\unskip}     \fi
\ifx \showDOI      \undefined \def \showDOI       #1{#1}\fi
\ifx \showISBNx    \undefined \def \showISBNx     #1{\unskip}     \fi
\ifx \showISBNxiii \undefined \def \showISBNxiii  #1{\unskip}     \fi
\ifx \showISSN     \undefined \def \showISSN      #1{\unskip}     \fi
\ifx \showLCCN     \undefined \def \showLCCN      #1{\unskip}     \fi
\ifx \shownote     \undefined \def \shownote      #1{#1}          \fi
\ifx \showarticletitle \undefined \def \showarticletitle #1{#1}   \fi
\ifx \showURL      \undefined \def \showURL       {\relax}        \fi
\providecommand\bibfield[2]{#2}
\providecommand\bibinfo[2]{#2}
\providecommand\natexlab[1]{#1}
\providecommand\showeprint[2][]{arXiv:#2}

\bibitem[\protect\citeauthoryear{??}{bok}{[n.d.]a}]%
        {bokehPy80:online}
 \bibinfo{year}{[n.d.]}\natexlab{a}.
\newblock \bibinfo{title}{bokeh · PyPI}.
\newblock \bibinfo{howpublished}{\url{https://pypi.org/project/bokeh/}}.
\newblock


\bibitem[\protect\citeauthoryear{??}{bok}{[n.d.]b}]%
        {bokehbok12:online}
 \bibinfo{year}{[n.d.]}\natexlab{b}.
\newblock \bibinfo{title}{bokeh/bokeh: Interactive Data Visualization in the
  browser, from Python}.
\newblock \bibinfo{howpublished}{\url{https://github.com/bokeh/bokeh}}.
\newblock


\bibitem[\protect\citeauthoryear{??}{bok}{[n.d.]c}]%
        {bokehbok61:online}
 \bibinfo{year}{[n.d.]}\natexlab{c}.
\newblock \bibinfo{title}{@bokeh/bokehjs - npm}.
\newblock
  \bibinfo{howpublished}{\url{https://www.npmjs.com/package/@bokeh/bokehjs}}.
\newblock


\bibitem[\protect\citeauthoryear{??}{For}{[n.d.]}]%
        {FortAwes12:online}
 \bibinfo{year}{[n.d.]}\natexlab{}.
\newblock \bibinfo{title}{FortAwesome/Font-Awesome: The iconic SVG, font, and
  CSS toolkit}.
\newblock
  \bibinfo{howpublished}{\url{https://github.com/fortawesome/Font-Awesome}}.
\newblock


\bibitem[\protect\citeauthoryear{??}{Git}{[n.d.]}]%
        {GitHubRE50:online}
 \bibinfo{year}{[n.d.]}\natexlab{}.
\newblock \bibinfo{title}{GitHub REST API documentation - GitHub Docs}.
\newblock
  \bibinfo{howpublished}{\url{https://docs.github.com/en/rest?apiVersion=2022-11-28}}.
\newblock


\bibitem[\protect\citeauthoryear{??}{Jav}{[n.d.]}]%
        {JavaOrac42:online}
 \bibinfo{year}{[n.d.]}\natexlab{}.
\newblock \bibinfo{title}{Java | Oracle}.
\newblock \bibinfo{howpublished}{\url{https://www.java.com/en/}}.
\newblock


\bibitem[\protect\citeauthoryear{??}{Lib}{[n.d.]}]%
        {Librarie69:online}
 \bibinfo{year}{[n.d.]}\natexlab{}.
\newblock \bibinfo{title}{Libraries.io - The Open Source Discovery Service}.
\newblock \bibinfo{howpublished}{\url{https://libraries.io/}}.
\newblock


\bibitem[\protect\citeauthoryear{??}{nei}{[n.d.]}]%
        {neilerns80:online}
 \bibinfo{year}{[n.d.]}\natexlab{}.
\newblock \bibinfo{title}{neilernst/cliffsDelta}.
\newblock
  \bibinfo{howpublished}{\url{https://github.com/neilernst/cliffsDelta}}.
\newblock


\bibitem[\protect\citeauthoryear{??}{npm}{[n.d.]}]%
        {npm27:online}
 \bibinfo{year}{[n.d.]}\natexlab{}.
\newblock \bibinfo{title}{npm}.
\newblock \bibinfo{howpublished}{\url{https://www.npmjs.com/}}.
\newblock


\bibitem[\protect\citeauthoryear{??}{PyS}{[n.d.]}]%
        {PySparkD87:online}
 \bibinfo{year}{[n.d.]}\natexlab{}.
\newblock \bibinfo{title}{PySpark Documentation — PySpark 3.3.2
  documentation}.
\newblock
  \bibinfo{howpublished}{\url{https://spark.apache.org/docs/latest/api/python/}}.
\newblock


\bibitem[\protect\citeauthoryear{??}{RTh}{[n.d.]}]%
        {RTheRPro33:online}
 \bibinfo{year}{[n.d.]}\natexlab{}.
\newblock \bibinfo{title}{R: The R Project for Statistical Computing}.
\newblock \bibinfo{howpublished}{\url{https://www.r-project.org/}}.
\newblock


\bibitem[\protect\citeauthoryear{??}{Rub}{[n.d.]}]%
        {RubyProg16:online}
 \bibinfo{year}{[n.d.]}\natexlab{}.
\newblock \bibinfo{title}{Ruby Programming Language}.
\newblock \bibinfo{howpublished}{\url{https://www.ruby-lang.org/en/}}.
\newblock


\bibitem[\protect\citeauthoryear{Alfadel, Costa, Shihab, and
  Mkhallalati}{Alfadel et~al\mbox{.}}{2021}]%
        {alfadel2021use}
\bibfield{author}{\bibinfo{person}{Mahmoud Alfadel},
  \bibinfo{person}{Diego~Elias Costa}, \bibinfo{person}{Emad Shihab}, {and}
  \bibinfo{person}{Mouafak Mkhallalati}.} \bibinfo{year}{2021}\natexlab{}.
\newblock \showarticletitle{On the Use of Dependabot Security Pull Requests}.
  In \bibinfo{booktitle}{\emph{2021 IEEE/ACM 18th International Conference on
  Mining Software Repositories (MSR)}}. \bibinfo{pages}{254--265}.
\newblock


\bibitem[\protect\citeauthoryear{Chen, Gao, and Xing}{Chen
  et~al\mbox{.}}{2016a}]%
        {analogicalqa}
\bibfield{author}{\bibinfo{person}{Chunyang Chen}, \bibinfo{person}{Sa Gao},
  {and} \bibinfo{person}{Zhenchang Xing}.} \bibinfo{year}{2016}\natexlab{a}.
\newblock \showarticletitle{Mining Analogical Libraries in Q amp;A Discussions
  -- Incorporating Relational and Categorical Knowledge into Word Embedding}.
  In \bibinfo{booktitle}{\emph{2016 IEEE 23rd International Conference on
  Software Analysis, Evolution, and Reengineering (SANER)}},
  Vol.~\bibinfo{volume}{1}. \bibinfo{pages}{338--348}.
\newblock
\urldef\tempurl%
\url{https://doi.org/10.1109/SANER.2016.21}
\showDOI{\tempurl}


\bibitem[\protect\citeauthoryear{Chen, Gao, and Xing}{Chen
  et~al\mbox{.}}{2016b}]%
        {Chen2016QA}
\bibfield{author}{\bibinfo{person}{Chunyang Chen}, \bibinfo{person}{Sa Gao},
  {and} \bibinfo{person}{Zhenchang Xing}.} \bibinfo{year}{2016}\natexlab{b}.
\newblock \showarticletitle{Mining Analogical Libraries in Q amp;A Discussions
  -- Incorporating Relational and Categorical Knowledge into Word Embedding}.
  In \bibinfo{booktitle}{\emph{2016 IEEE 23rd International Conference on
  Software Analysis, Evolution, and Reengineering (SANER)}},
  Vol.~\bibinfo{volume}{1}. \bibinfo{pages}{338--348}.
\newblock
\urldef\tempurl%
\url{https://doi.org/10.1109/SANER.2016.21}
\showDOI{\tempurl}


\bibitem[\protect\citeauthoryear{Cohen}{Cohen}{2013}]%
        {cohen2013statistical}
\bibfield{author}{\bibinfo{person}{Jacob Cohen}.}
  \bibinfo{year}{2013}\natexlab{}.
\newblock \bibinfo{booktitle}{\emph{Statistical power analysis for the
  behavioral sciences}}.
\newblock \bibinfo{publisher}{Routledge}.
\newblock


\bibitem[\protect\citeauthoryear{Constantinou, Decan, and Mens}{Constantinou
  et~al\mbox{.}}{2018}]%
        {Constantinou2018BreakingTB}
\bibfield{author}{\bibinfo{person}{Eleni Constantinou},
  \bibinfo{person}{Alexandre Decan}, {and} \bibinfo{person}{Tom Mens}.}
  \bibinfo{year}{2018}\natexlab{}.
\newblock \showarticletitle{Breaking the Borders: An Investigation of
  Cross-Ecosystem Software Packages}. In
  \bibinfo{booktitle}{\emph{BElgian-NEtherlands software eVOLution symposium}}.
\newblock


\bibitem[\protect\citeauthoryear{Cossette and Walker}{Cossette and
  Walker}{2012}]%
        {Cossette2021recommendation}
\bibfield{author}{\bibinfo{person}{Bradley~E. Cossette} {and}
  \bibinfo{person}{Robert~J. Walker}.} \bibinfo{year}{2012}\natexlab{}.
\newblock \showarticletitle{Seeking the Ground Truth: A Retroactive Study on
  the Evolution and Migration of Software Libraries}. In
  \bibinfo{booktitle}{\emph{Proceedings of the ACM SIGSOFT 20th International
  Symposium on the Foundations of Software Engineering}}. Article
  \bibinfo{articleno}{55}, \bibinfo{numpages}{11}~pages.
\newblock


\bibitem[\protect\citeauthoryear{César, Harrand, Monperrus, and Baudry}{César
  et~al\mbox{.}}{2021}]%
        {Cesar:emse2021}
\bibfield{author}{\bibinfo{person}{Soto-Valero César},
  \bibinfo{person}{Nicolas Harrand}, \bibinfo{person}{Martin Monperrus}, {and}
  \bibinfo{person}{Benoit Baudry}.} \bibinfo{year}{2021}\natexlab{}.
\newblock \showarticletitle{A Comprehensive Study of Bloated Dependencies in
  the Maven Ecosystem}.
\newblock \bibinfo{journal}{\emph{Empirical Software Engineering}}
  \bibinfo{volume}{26}, \bibinfo{number}{3} (\bibinfo{year}{2021}).
\newblock


\bibitem[\protect\citeauthoryear{Decan, Mens, and Claes}{Decan
  et~al\mbox{.}}{2016}]%
        {10.1145/2993412.3003382}
\bibfield{author}{\bibinfo{person}{Alexandre Decan}, \bibinfo{person}{Tom
  Mens}, {and} \bibinfo{person}{Maelick Claes}.}
  \bibinfo{year}{2016}\natexlab{}.
\newblock \showarticletitle{On the Topology of Package Dependency Networks: A
  Comparison of Three Programming Language Ecosystems}. In
  \bibinfo{booktitle}{\emph{Proccedings of the 10th European Conference on
  Software Architecture Workshops}}. Article \bibinfo{articleno}{21},
  \bibinfo{numpages}{4}~pages.
\newblock
\urldef\tempurl%
\url{https://doi.org/10.1145/2993412.3003382}
\showDOI{\tempurl}


\bibitem[\protect\citeauthoryear{Decan, Mens, and Grosjean}{Decan
  et~al\mbox{.}}{2019}]%
        {decan2019empirical}
\bibfield{author}{\bibinfo{person}{Alexandre Decan}, \bibinfo{person}{Tom
  Mens}, {and} \bibinfo{person}{Philippe Grosjean}.}
  \bibinfo{year}{2019}\natexlab{}.
\newblock \showarticletitle{An empirical comparison of dependency network
  evolution in seven software packaging ecosystems}.
\newblock \bibinfo{journal}{\emph{Empirical Software Engineering}}
  \bibinfo{volume}{24}, \bibinfo{number}{1} (\bibinfo{year}{2019}),
  \bibinfo{pages}{381--416}.
\newblock


\bibitem[\protect\citeauthoryear{Islam, Kula, Treude, Chinthanet, Ishio, and
  Matsumoto}{Islam et~al\mbox{.}}{2021}]%
        {Syful2021ICSME}
\bibfield{author}{\bibinfo{person}{Syful Islam},
  \bibinfo{person}{Raula~Gaikovina Kula}, \bibinfo{person}{Christoph Treude},
  \bibinfo{person}{Bodin Chinthanet}, \bibinfo{person}{Takashi Ishio}, {and}
  \bibinfo{person}{Kenichi Matsumoto}.} \bibinfo{year}{2021}\natexlab{}.
\newblock \showarticletitle{Contrasting Third-Party Package Management User
  Experience}. In \bibinfo{booktitle}{\emph{2021 IEEE International Conference
  on Software Maintenance and Evolution (ICSME)}}. \bibinfo{pages}{664--668}.
\newblock
\urldef\tempurl%
\url{https://doi.org/10.1109/ICSME52107.2021.00077}
\showDOI{\tempurl}


\bibitem[\protect\citeauthoryear{Katz}{Katz}{2020}]%
        {jeremy_katz_2020_3626071}
\bibfield{author}{\bibinfo{person}{Jeremy Katz}.}
  \bibinfo{year}{2020}\natexlab{}.
\newblock \bibinfo{booktitle}{\emph{{Libraries.io Open Source Repository and
  Dependency Metadata}}}.
\newblock
\urldef\tempurl%
\url{https://doi.org/10.5281/zenodo.3626071}
\showDOI{\tempurl}


\bibitem[\protect\citeauthoryear{Kula, German, Ouni, Ishio, and Inoue}{Kula
  et~al\mbox{.}}{2018}]%
        {KulaEMSE2018}
\bibfield{author}{\bibinfo{person}{Raula~Gaikovina Kula},
  \bibinfo{person}{Daniel~M. German}, \bibinfo{person}{Ali Ouni},
  \bibinfo{person}{Takashi Ishio}, {and} \bibinfo{person}{Katsuro Inoue}.}
  \bibinfo{year}{2018}\natexlab{}.
\newblock \showarticletitle{Do Developers Update Their Library Dependencies?}
\newblock \bibinfo{journal}{\emph{Empirical Software Engineering}}
  \bibinfo{volume}{23}, \bibinfo{number}{1} (\bibinfo{year}{2018}),
  \bibinfo{pages}{384–417}.
\newblock
\urldef\tempurl%
\url{https://doi.org/10.1007/s10664-017-9521-5}
\showDOI{\tempurl}


\bibitem[\protect\citeauthoryear{Mann and Whitney}{Mann and Whitney}{1947}]%
        {mann1947test}
\bibfield{author}{\bibinfo{person}{Henry~B Mann} {and}
  \bibinfo{person}{Donald~R Whitney}.} \bibinfo{year}{1947}\natexlab{}.
\newblock \showarticletitle{On a test of whether one of two random variables is
  stochastically larger than the other}.
\newblock \bibinfo{journal}{\emph{The annals of mathematical statistics}}
  (\bibinfo{year}{1947}), \bibinfo{pages}{50--60}.
\newblock


\bibitem[\protect\citeauthoryear{McNemar}{McNemar}{1947}]%
        {McNemar1947}
\bibfield{author}{\bibinfo{person}{Quinn McNemar}.}
  \bibinfo{year}{1947}\natexlab{}.
\newblock \showarticletitle{Note on the sampling error of the difference
  between correlated proportions or percentages}.
\newblock \bibinfo{journal}{\emph{Psychometrika}} \bibinfo{volume}{12},
  \bibinfo{number}{2} (\bibinfo{date}{June} \bibinfo{year}{1947}),
  \bibinfo{pages}{153--157}.
\newblock
\urldef\tempurl%
\url{https://doi.org/10.1007/bf02295996}
\showDOI{\tempurl}


\bibitem[\protect\citeauthoryear{Romano, Kromrey, Coraggio, Skowronek, and
  Devine}{Romano et~al\mbox{.}}{2006}]%
        {romano2006exploring}
\bibfield{author}{\bibinfo{person}{Jeanine Romano}, \bibinfo{person}{Jeffrey~D
  Kromrey}, \bibinfo{person}{Jesse Coraggio}, \bibinfo{person}{Jeff Skowronek},
  {and} \bibinfo{person}{Linda Devine}.} \bibinfo{year}{2006}\natexlab{}.
\newblock \showarticletitle{Exploring methods for evaluating group differences
  on the NSSE and other surveys: Are the t-test and Cohen’sd indices the most
  appropriate choices}. In \bibinfo{booktitle}{\emph{annual meeting of the
  Southern Association for Institutional Research}}. \bibinfo{pages}{1--51}.
\newblock


\bibitem[\protect\citeauthoryear{Teyton, Falleri, and Blanc}{Teyton
  et~al\mbox{.}}{2013}]%
        {Teyton2013map}
\bibfield{author}{\bibinfo{person}{Cédric Teyton}, \bibinfo{person}{Jean-Rémy
  Falleri}, {and} \bibinfo{person}{Xavier Blanc}.}
  \bibinfo{year}{2013}\natexlab{}.
\newblock \showarticletitle{Automatic discovery of function mappings between
  similar libraries}. In \bibinfo{booktitle}{\emph{2013 20th Working Conference
  on Reverse Engineering (WCRE)}}. \bibinfo{pages}{192--201}.
\newblock
\urldef\tempurl%
\url{https://doi.org/10.1109/WCRE.2013.6671294}
\showDOI{\tempurl}


\bibitem[\protect\citeauthoryear{Valiev, Vasilescu, and Herbsleb}{Valiev
  et~al\mbox{.}}{2018}]%
        {fse2018_sustained}
\bibfield{author}{\bibinfo{person}{Marat Valiev}, \bibinfo{person}{Bogdan
  Vasilescu}, {and} \bibinfo{person}{James Herbsleb}.}
  \bibinfo{year}{2018}\natexlab{}.
\newblock \showarticletitle{Ecosystem-Level Determinants of Sustained Activity
  in Open-Source Projects: A Case Study of the PyPI Ecosystem}. In
  \bibinfo{booktitle}{\emph{Proceedings of the 2018 26th ACM Joint Meeting on
  European Software Engineering Conference and Symposium on the Foundations of
  Software Engineering}}. \bibinfo{pages}{644–655}.
\newblock


\bibitem[\protect\citeauthoryear{Wilcoxon}{Wilcoxon}{1945}]%
        {wilcoxon1945individual}
\bibfield{author}{\bibinfo{person}{Frank Wilcoxon}.}
  \bibinfo{year}{1945}\natexlab{}.
\newblock \bibinfo{title}{Individual comparisons by ranking methods. Biometrics
  Bulletin, 1 (6), 80-83}.
\newblock
\newblock


\bibitem[\protect\citeauthoryear{Zerouali, Mens, Robles, and
  Gonzalez-Barahona}{Zerouali et~al\mbox{.}}{2019}]%
        {zerouali2019diversity}
\bibfield{author}{\bibinfo{person}{Ahmed Zerouali}, \bibinfo{person}{Tom Mens},
  \bibinfo{person}{Gregorio Robles}, {and} \bibinfo{person}{Jesus~M
  Gonzalez-Barahona}.} \bibinfo{year}{2019}\natexlab{}.
\newblock \showarticletitle{On the diversity of software package popularity
  metrics: An empirical study of npm}. In \bibinfo{booktitle}{\emph{2019 IEEE
  26th International Conference on Software Analysis, Evolution and
  Reengineering (SANER)}}. \bibinfo{pages}{589--593}.
\newblock


\end{thebibliography}
\balance
\end{sloppy}
\end{document}